\documentclass[prl,twocolumn,amsmath,amssymb, aps]{revtex4-1}

\usepackage{graphicx}
\usepackage{dcolumn}
\usepackage{bm}
\usepackage{lipsum}
\usepackage{nameref}
\usepackage{varioref}
\usepackage{hyperref}
\usepackage{cleveref}
\newcommand{\minus}{\scalebox{0.75}[1.0]{$-$}}
\usepackage{float}
\usepackage[export]{adjustbox}
\usepackage{subscript}
\usepackage{color}
\usepackage[normalem]{ulem}

\newcommand{\sco}{Sr\textsubscript{2}CuO\textsubscript{3}}
\begin{document}
 \preprint{APS/123-QED}
 \title{Multi-spinon and holon excitations probed by resonant inelastic x-ray scattering 
  on doped one-dimensional antiferromagnets}
 \author{Umesh Kumar$^{1,2}$, Alberto Nocera$^{1,3}$,  Elbio Dagotto$^{1,3}$, and Steven Johnston$^{1,2}$}
 \affiliation{$^1$Department of Physics and Astronomy, The University of Tennessee, Knoxville, TN 37996, USA \\
  $^2$Joint Institute for Advanced Materials, The University of Tennessee, Knoxville, TN 37996, USA \\ 
  $^3$Materials Science and Technology Division, Oak Ridge National Laboratory, Oak Ridge, Tennessee 37831, USA}
 \date{\today}
 \begin{abstract}
  Resonant inelastic x-ray scattering (RIXS) at the
  oxygen $K$-edge has recently accessed multi-spinon excitations in the
one-dimensional antiferromagnet (1D-AFM) \sco, where four-spinon excitations
are resolved separately from the two-spinon continuum. This technique,
therefore,  provides new opportunities to study fractionalized quasiparticle
excitations in doped 1D-AFMs. To this end, we carried out exact diagonalization
studies of the doped $t$-$J$ model and provided predictions for oxygen $K$-edge
RIXS experiments on doped 1D-AFMs. We show that the RIXS spectra are rich,
containing distinct two- and four-spinon excitations, dispersive (anti)holon
excitations, and combinations thereof. Our results highlight how RIXS
complements inelastic neutron scattering experiments by accessing additional 
charge and spin components of fractionalized quasiparticles.
 \end{abstract}    
 \pacs{Valid PACS appear here}
 \maketitle

 \textit{Introduction} --- One-dimensional (1D) magnetic systems have attracted considerable interest throughout the scientific community for more than half a century. 
 This interest stems from the fact that these systems provide excellent opportunities to study novel quantum phenomena such as quasiparticle fractionalization or quantum criticality. 
 Moreover, model Hamiltonians of 1D systems can often be solved 
 exactly using analytical or numerical techniques, 
making them ideal starting points for understanding the physics of strongly correlated materials.
 For example, the exact solution of the 1D Hubbard model by Lieb and 
 Wu~\cite{PhysRevLett.20.1445} represented a breakthrough in the field, showing that interacting electrons confined to 1D are characterized by spin-charge separation, 
where electronic quasiparticle excitations \emph{break} into collective density fluctuations carrying either spinless charge (``holons'') or chargeless spin (``spinons'') quantum numbers with different characteristic energy scales. This work inspired an intense search for materials showing spin-charge separation, but it has only been in the last two decades that this phenomenon was observed \cite{PhysRevLett.77.4054,Segovia1999,PhysRevB.59.7358,Kim2006,PhysRevB.73.201101,Jompol597}.

\begin{figure}[h] 
  \centering
  \includegraphics[width=\linewidth,center]{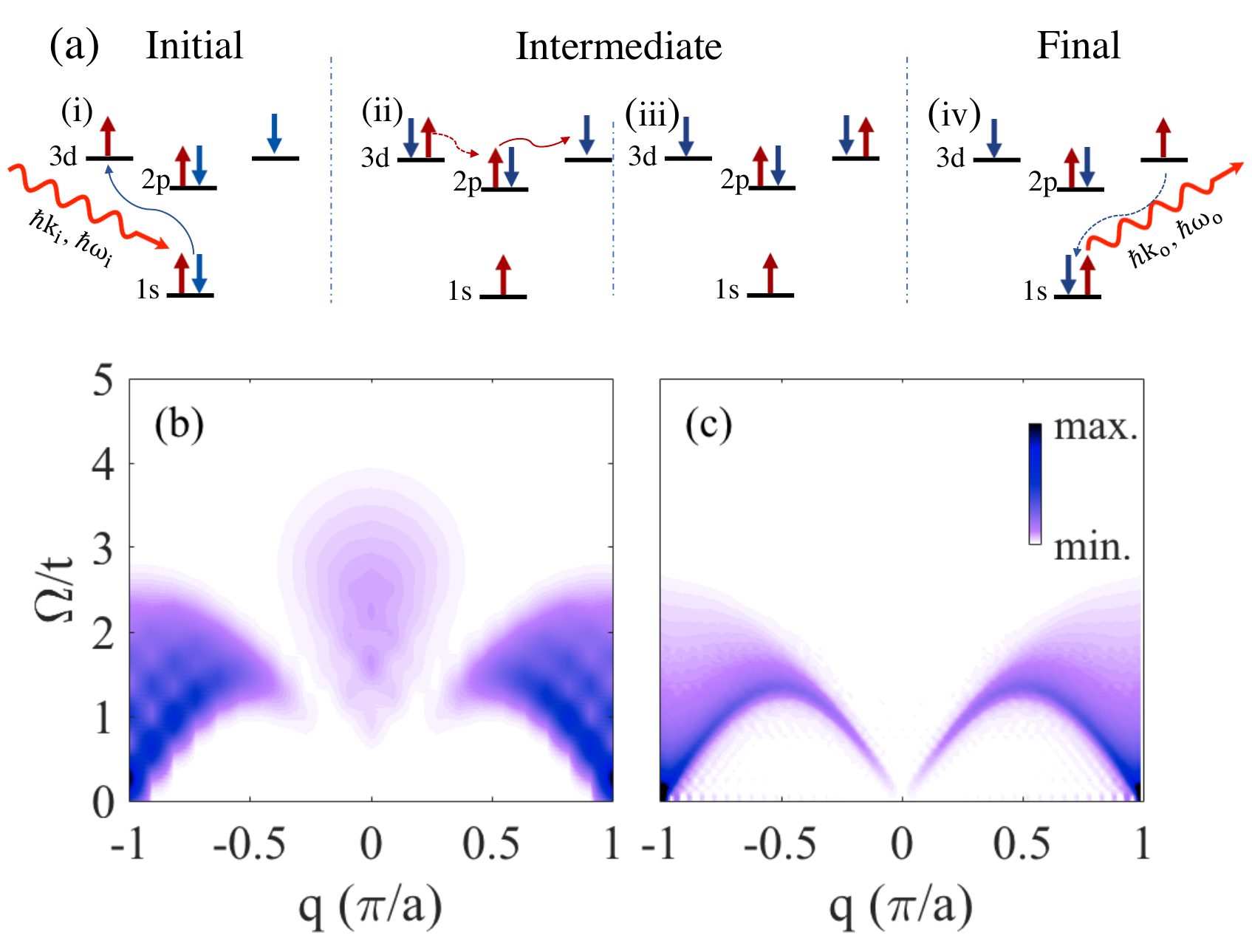}
  \caption{\label{spinflip} a) A sketch of the spin-flip mechanism in oxygen
$K$-edge RIXS. Hybridization between the Cu and O orbitals allows an 
incident photon to excite an O $1s$ electron into the $3d_{x^2-y^2}$ orbital
on one of the two neighboring Cu sites, creating a Cu d$^{10}$ upper Hubbard
band excitation in the intermediate state (subpanel i). The $d^{10}$
excitation can transfer to the other neighboring Cu site via two 
Cu-O hopping processes [(subpanel ii) \& (subpanel iii)]. Finally, the extra electron decays back
into the O $1s$ core level, leaving the system in a final state with a double
spin-flip (subpanel iv). b) Computed RIXS spectra for an undoped $t$-$J$ model on a 22-site 
chain. c) $S(q, \omega)$ for an undoped 80-site chain, calculated with DMRG for 
the same model. Note the additional spectral weight in the RIXS intensity 
centered at $q = 0$, and absent in $S(q,\omega)$.}
 \vskip -0.15cm
 \end{figure}

 Resonant inelastic x-ray scattering (RIXS) \cite{RevModPhys.83.705} has evolved as an important tool for studying the magnetic excitations in correlated materials~\cite{PhysRevLett.100.097001,PhysRevLett.110.147001,
        PhysRevLett.103.117003}, complementing inelastic neutron scattering (INS). 
        RIXS, however, is also a powerful probe of orbital and charge 
        excitations, as was succinctly demonstrated by the experimental observation 
        of spin-orbital fractionalization in a Cu $L$-edge RIXS study of 
        \sco~\cite{Nature.485.7396,PhysRevB.88.195138}. 
 \sco~ contains 1D chains of corner-shared CuO$_4$ plaquettes, where a single hole 
        occupies each Cu $3d_{x^2-y^2}$ orbital, forming a quasi-1D spin-$\frac{1}{2}$ chain.  
        Due to a very weak interchain interaction, 
        the CuO$_3$ chains decouple above the bulk ordering temperature $T_N = 5.5$ K and form a nearly ideal realization of a 1D antiferromagnet (AFM) \cite{Kojima1997}. 
        A recent O $K$-edge RIXS study~\cite{FSDIR} of undoped \sco~ 
  directly observed multi-spinon excitations outside of the two-spinon (2S) continuum 
        (see also Fig. \ref{spinflip}) further highlighting the potential for RIXS to probe such excitations. 
       
        To date, spin-charge separation has not been observed using 
 RIXS~\cite{PhysRevB.88.195138}. In this letter, we performed exact diagonalization (ED) and DMRG calculations to show that RIXS measurements on doped 1D AFMs can fill this need. Specifically, we show that O $K$-edge RIXS can access multi-spinon excitations, antiholon excitations, and combinations thereof, thus providing a unique view of spin-charge separation in doped 1D AFMs. Since \sco~can be doped with Zn, Ni, or  Co~\cite{doi:10.1021/acs.cgd.5b00652, PhysRevB.95.235154}, this material can be used to test our predictions.  
  
{\it Magnetic Scattering at the O $K$-edge} --- Before proceeding, we review how magnetic excitations occur in the O $K$-edge ($1s\rightarrow 2p$)~\cite{LuukThesis} measurements on \sco, as sketched in Fig. \ref{spinflip}(a).          
        \sco~is a charge-transfer insulator and the ground state character of the 
        CuO$_4$ plaquettes is predominantly of the form 
 $\alpha|d^9\rangle + \beta |d^{10}\underbar{L}\rangle$ 
 ($\alpha^2\approx0.64$, $\beta^2 \approx 0.36$) \cite{PhysRevB.62.10752, Walters2009}, due to hybridization between the Cu 3$d_{x^2-y^2}$ and O 2$p$ orbitals. Here, $\underbar{L}$ denotes a hole on the ligand O orbitals. Due to this hybridization, the incident photon can excite an O $1s$ core electron into the Cu $3d$ orbital when tuned to the O $K$-edge, creating an upper Hubbard band excitation. In the intermediate state, the $d^{10}$ configuration can move to the neighboring Cu ion via the bridging O orbital. Since the adjacent Cu orbital also hybridizes with the O containing the core hole, one of the $d^{10}$ electrons can then decay to fill it, creating a final state with a double spin flip. 

The dynamics in the intermediate state are essential for generating magnetic excitations at this edge, and this is a fundamental difference in how RIXS and INS probe magnetic excitations. One of the advantages of working at the O $K$-edge is that it has 
 relatively long core-hole lifetimes ($\hbar/\Gamma$,~$\Gamma$ = 0.15 eV \cite{PhysRevLett.110.265502}) in comparison to other edges ($\Gamma = 1.5$~eV at the Cu $K$-edge and $0.3$~eV at the Cu L\textsubscript{3}-edge \cite{PhysRevB.85.064422}), which provides a longer window for generating magnetic excitations \cite{FSDIR,TohyamaPreprint}. 
Because of this, the inclusion of the intermediate states in the modeling is necessary. Several efforts addressing the spin dynamics in RIXS have mostly used the {\it ultrashort core-hole lifetime} (UCL) approximations, which applies to edges with short core-hole lifetimes \cite{PhysRevB.75.115118, PhysRevLett.106.157205}, while studies of 1D systems beyond UCL approximations have been limited~\cite{PhysRevB.85.064423, PhysRevB.85.064422}. Ref. \onlinecite{PhysRevB.85.064423} studied the effect of incidence energy on spin dynamics RIXS spectra in 1D using small cluster ED, but a systematic analysis of the incident energy dependence was not carried out. As a result, the  
 multi-spinon excitations at $q = 0$ were not reported. Similarly, Ref.~\onlinecite{PhysRevB.83.245133} discussed the doping dependence of the RIXS spectrum for the $t$-$J$ model  
by evaluating the spin response, but the charge response along with the intermediate state dynamics were left out.
For these reasons, these prior studies could not address the physics reported here. 

 \textit{Model and methods} ---  
 We model the RIXS spectra of \sco~using a 1D $t$-$J$ Hamiltonian  
 \begin{equation*}
 H = -t \sum_{ i , \sigma}(\tilde{d}_{ i, \sigma }^\dagger 
 \tilde{d}_{i+1,\sigma}^{\phantom\dagger}+h.c.) 
 + J\sum_{ i }(\textbf{S}_i\cdot \textbf{S}_{i+1} - \frac{1}{4}n_i n_{i+1}).
 \end{equation*}
 Here, $\tilde{d}_{i,\sigma }$ is the annihilation operator for a hole with spin
 $\sigma$ at site $i$, under the constraint of no double occupancy, $n_i = 
 \sum_\sigma n_{i,\sigma}$ is the number operator, and $\textbf{S}_i$ is the
 spin operator at site $i$. 
 
During the RIXS process~\cite{RevModPhys.83.705}, an incident photon with
momentum ${\bf k}_\mathrm{in}$ and energy $\omega_\mathrm{in}$ ($\hbar = 1$) 
tuned to an elemental absorption edge resonantly excites a core electron into
an unoccupied state in the sample. The resulting core hole and excited electron
interact with the system creating several elementary excitations before an 
electron radiatively decays into the core level, emitting a photon with
energy $\omega_\mathrm{out}$ and momentum ${\bf k}_\mathrm{out}$. The RIXS
intensity is given by the Kramers-Heisenberg formula~\cite{RevModPhys.83.705}   
 \begin{equation}\label{eq:KH}
 I = \sum_f \bigg|\sum_{n} \frac{\langle  f| D^{\dagger } |n \rangle \langle n | D | i \rangle }{ E_i + \omega_\mathrm{in} - E_n + i \Gamma_n} \bigg|^2\delta(E_f-E_i-\Omega),  
 \end{equation} 
 where $\Omega$ = $\omega_\mathrm{in}-\omega_\mathrm{out}$ is the energy loss,  
 $|i\rangle$, $|n\rangle$, and $|f\rangle$ are the initial, intermediate, and final states 
 of the RIXS process with energies $E_i$, $E_n$, and $E_f$, respectively, and $D$ is the dipole 
 operator for the O $1s \rightarrow 2p$ transition.  
 In the downfolded $t$-$J$ model $D$ takes the effective form 
 \begin{equation*}
 D = \sum_{i,\sigma} e^{i\bf{k}_\mathrm{in}\cdot(\bf{R_i} + \bf{a}/2)}
 \left[\left(\tilde{d}_{i, \sigma }^{\phantom{\dagger}} \minus \tilde{d}_{i+1,\sigma}^{\phantom{\dagger}}\right)
 s_{i+\frac{1}{2}, \sigma }^{\dagger}+ h.c.\right],  
 \end{equation*}
 where ${\bf q}\ (= {\bf k}_{\mathrm{out}}-{\bf k}_\mathrm{in})$ is the
momentum transfer and the relative sign is due to the phases of the Cu
$3d_{x^2-y^2}$ and O $2p_x$ orbital overlaps along the chain direction. Here,
$s_{i+\frac{1}{2},\sigma}$ is the hole annihilation operator for the $1s$ core
level on the O atom bridging the $i$ and $i+1$ Cu sites. 
 
 In the real material, the core hole potential raises the on-site energy of the
bridging oxygen orbital (in hole language) in the intermediate state while
exerting a minimal influence on the Cu sites. This change locally modifies the
superexchange interaction between the neighboring Cu atoms
\cite{PhysRevB.48.9788}. To account for this effect, we reduce the value of
$J_{i,i+1} = J/2$ when solving for the intermediate states, where the core-hole
is created on the O atom bridging the $i$ and $i+1$ sites. Our results are not
sensitive to reasonable changes in this value \cite{Supplement}.  
 
 Throughout we set $t = 1$ as our unit of energy ($t \approx 300$ meV in \sco).
The remaining parameters are  $\Gamma_n = \frac{1}{2}t$ for all $n$ and $J$ =
$\frac{5}{6}t$, unless otherwise stated. These values are typical for the O
$K$-edge measurements of Sr$_2$CuO$_3$
\cite{Nature.485.7396,PhysRevB.85.214527, PhysRevLett.110.265502, Walters2009}. 
 We also introduce a Gaussian broadening ($\Gamma= \frac{1}{3}t$) for energy conserving
$\delta$-function appearing in Eq. (\ref{eq:KH}). We evaluated Eq.
(\ref{eq:KH}) on a $L=20$ site chain using the Lanczos method with a fixed filling. 
 
 To help identify the relevant charge and spin excitations in the RIXS spectra,
we also performed DMRG simulations~\cite{White92,White93} for the dynamical
charge $N(q,\omega)$ and spin $S(q,\omega)$ structure factors on an $L = 80$
site chain, and using correction-vector
method~\cite{Kuhner1999,Jeckelmann2002}. Within the correction vector approach,
we used the Krylov decomposition~\cite{nocera2016spectral} instead of the
conjugate gradient. In the ground state and dynamic DMRG simulations, 
 we used a maximum of $m=1000$ states, keeping the truncation error below
$10^{-6}$ and used a broadening of the correction-vector calculation of $\eta=0.08t$ 
\cite{Supplement}. The computer 
 package \textsc{dmrg++} developed by G. Alvarez, CNMS, ORNL, 
 was used in the DMRG simulations \cite{alvarez09}.
\\

 \begin{figure}[t] 
  \includegraphics[width=0.975\columnwidth, angle =-90,center]{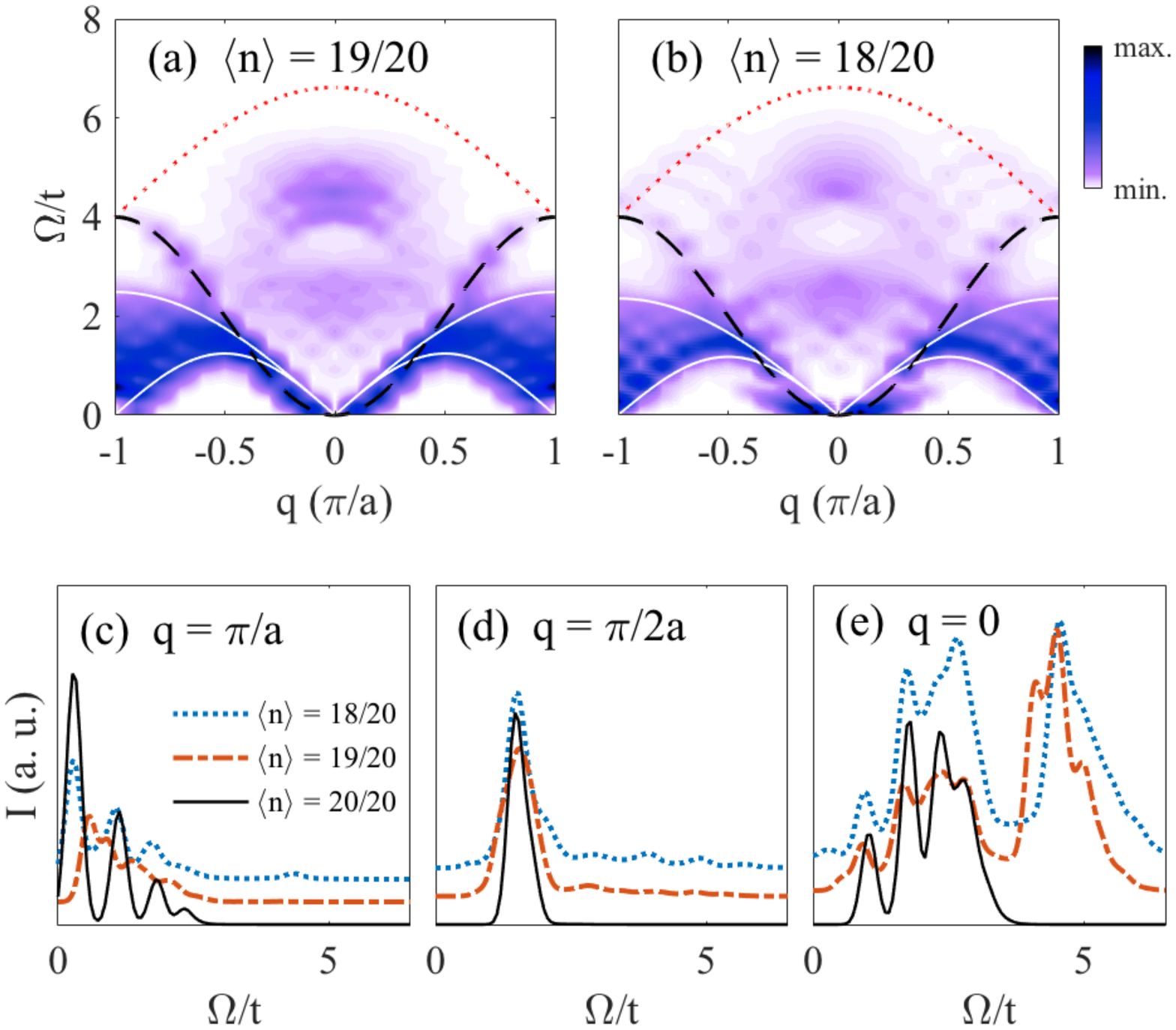}
  \vskip -0.5cm
  \caption{\label{dn1n2} Calculated RIXS spectra at $\omega_\mathrm{in} = 5t$ for a $L = 20$ site doped $t$-$J$ chain with (a) one and (b) two additional doped electrons. The white, black, and red lines shows the boundaries for two-spinon continuum, the dispersion of the antiholonic excitation, and upper boundary for antiholon-2S excitations, respectively. Panels (c), (d), and (e) compare the doped RIXS spectra with the undoped case 
for momentum transfers of $q = \pi/a$, $q = \pi/(2a)$, and $q = 0$, respectively.}
  \vskip -0.5cm
 \end{figure}
 
 \textit{Undoped RIXS spectra} --- Figure \ref{spinflip}(b) shows the RIXS 
 intensity for the half-filled $t$-$J$ chain, reproduced from Ref.~\onlinecite{FSDIR}. 
 For comparison, Fig.~\ref{spinflip}(c) shows $S(q,\omega)$ 
 obtained using DMRG for the same parameters. 
 The RIXS intensity has two main features. The first is a  
 continuum of excitations that closely mirrors $S(q,\omega)$ and is situated within the boundaries of the 2S continuum. 
Its intensity is relatively independent of the incident photon energy and is associated primarily with 2S excitations~\cite{FSDIR, FOUR}. 
 The second feature is a continuum of excitations laying outside of the 2S continuum, corresponding to 4S excitations. Its intensity is sensitive to both the incident photon energy and the core-hole lifetime, indicating that the intermediate state plays a critical role in creating those excitations~\cite{FSDIR}.   
 
 {\textit{Doped RIXS spectra ---} We now turn our attention to the results for the doped case. Figures \ref{dn1n2}(a) and \ref{dn1n2}(b) show the RIXS intensity obtained on a 20-site chain 
 at 5\% and 10\% electron doping, respectively. Here, we have used $\omega_\mathrm{in} = 5t$ 
 to enhance the intensity of the features appearing at $q = 0$. 
 To help us better understand the main features, we also computed $S(q,\omega)$ (Fig.~\ref{Fig:DMRG}(a)) and $N(q,\omega)$ (Fig.~\ref{Fig:DMRG}(b)) for 5\% doping using DMRG. 
 
\begin{figure}[t]
\vskip -1.0cm
\includegraphics[width=0.8\linewidth, angle =-90,center]{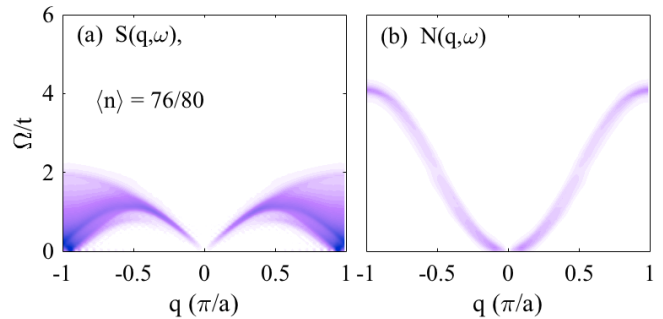}
\vskip -1.25cm
\caption{\label{Fig:DMRG} DMRG results for (a) $S(q, \omega)$ and (b) $N(q,\omega)$ for the doped $t$-$J$ model on a L = 80 sites chain and $\langle n \rangle$ = 0.95 doping.}
\label{dopedDMRG}
\vskip -0.4cm
\end{figure} 
The RIXS spectra for the doped cases have three recognizable sets of features: i) a continuum that mirrors the $S(q,\omega)$ in Fig.~\ref{Fig:DMRG}(a); ii) a cosine-like dispersive feature with a bandwidth of 4$t$ that mirrors $N(q,\omega)$ in Fig.~\ref{Fig:DMRG}(b); and iii) two continua, centered at $q=0$ and extending up to $\sim6t$ in energy loss. These features are absent in $S(q,\omega)$ and $N(q,\omega)$. The excitations (i) and (ii) point to a manifestation of spin-charge separation in that the response \emph{bifurcates} into primarily two-spinon (i) and antiholon (ii) excitations, characterized by different energy scales. 
Also, notice that the dispersions of various peaks in Figs. \ref{dn1n2}(a) and \ref{dn1n2}(b) do not vary significantly with a small change in doping, except for their relative intensities.   
    
 Figures \ref{dn1n2}(c)-\ref{dn1n2}(e) compare the doping evolution of the RIXS features at fixed momentum points. 
Fig. \ref{dn1n2}(c) shows $q = \pi/a$, where the upper bound 
($\pi J$) of the spin excitations decreases upon doping. 
Similarly, the line-cut at $q = \pi/2a$ in \ref{dn1n2}(d) shows that the lower bound ($\pi J/2$) of the 2S 
continuum also decreases with doping, allowing for final states below the 2S continuum of the undoped case. 
We also observe a secondary feature at higher energy loss due to changes in the 
holon branch and 4S excitations. Fig.\ref{dn1n2}(e) shows a 
 cut at $q = 0$, where two distinct sets of peaks are clear. 
The group at lower energy losses appears in the same energy range of the multi-spinon peak observed in the undoped case. The peaks at higher energy loss 
appear above $\Omega = 4t$ and are identified below. 
  
The calculated spectra can be understood by making use of the spin-charge separation picture: in 1D, the wavefunction of the large $U$ Hubbard model for $N$ electrons in $L$ lattice sites is a product of `spinless' charge and `chargeless' spin wavefunctions.~\cite{0022-3719-15-1-007,PhysRevB.41.2326, PhysRevB.55.15475} 
The dispersion of charge excitations is given
by $\omega_{\bar{h}}(k_{\bar{h}})=-2t\cos(k_{\bar{h}}a)$~\cite{0953-8984-4-13-020,KimPRB}, which agrees well with the 
dispersion observed for feature (ii) (see black dashed line) 
and in $N(q,\omega)$. As shown in the supplemental material~\cite{Supplement},  
the $N(q,\omega)$ computed here for small \emph{electron doping} 
is identical to the $N(q,\omega)$ obtained for a 1D spinless fermions chain 
with the same \emph{fermionic filling}, supporting the spin-charge 
separation picture. This result indicates that the charge excitation is behaving like a 
nearly free spinless quasiparticle, i.e. a holon/antiholon. 
Concerning the spin part, the dispersion relation for a single spinon is 
given by $\omega_s(k_s)=\frac{\pi}{2}J\vert \sin(k_s a)\vert$. 
Due to the RIXS selection rules, these spin excitations must be generated in even numbers, resulting in a continuum whose boundaries are defined by this dispersion relation. At small doping, the limits of this continuum are modified, which is 
accounted for using a slightly modified superexchange  
$\tilde{J} = J\langle n\rangle$~\cite{PhysRevB.55.15475}. 
The upper and lower boundaries of the modified 2S continuum are indicated by the white lines in Fig. \ref{dn1n2} and agree well with the observed excitations.   

We can summarize the picture emerging from our results as follows: the   
2S-like continuum present in the RIXS spectrum is a pure magnetic excitation 
as it compares well with the $S(q,\omega)$ from DMRG. 
The dispersing cosine-like feature in the doped RIXS
spectra compares well with the $N(q,\omega)$ from DMRG. We have verified that
the $N(q,\omega)$ of the spinless fermions with occupations equal to the
electron-doping considered above are qualitatively similar to the results obtained for the
doped $t$-$J$ chain \cite{Supplement}.  We therefore assign this feature
to purely charge-like antiholon excitations. 
 
The peaks at $q=0$ of the RIXS spectrum are not captured by either
$S(q,\omega)$ or $N(q,\omega)$. The lower continuum resembles the multi-spinon
continuum \cite{FSDIR} also observed in the undoped case, and we, therefore,
associate it with 4S excitations. Conversely, the continuum of excitations at
energy losses between $4t$ and $6.5t$ (well beyond the upper boundary of 4S
continuum [$2\pi J \ ( =5.24t)$]~\cite{1742-5468-2006-12-P12013}) is
unique to the doped case. The excitations are bounded by $4t+\pi J\cos(q/2)$
(dotted red line), which one obtains from a simple convolution of the antiholon
and two-spinon excitations. 
Therefore, we assign these to an antiholon plus two-spinon final state. The
fact that the intensity and distribution of these excitations are very
sensitive to doping supports this view. As we further increase the doping, we
see additional spectral weight above the $4t+\pi J\cos(q/2)$ boundary,
indicating that these quasiparticle interactions are beginning to interact to
produce modified dispersion relationships. 

\textit{Incidence energy dependence ---} Figure~\ref{Fig:incdep} shows the
changes in the RIXS intensity maps as the incident photon energy is varied from
$\omega_\mathrm{in} = 3t$ 
to $8t$ for the 5\% doped case. (The results at 10\% doping are similar and
provided in Ref. \onlinecite{Supplement}.) The final state excitations
resembling $S(q,\omega)$ and $N(q,\omega)$ are clear in all cases, 
but there are some variations in the overall intensity as $\omega_\mathrm{in}$
is tuned through the XAS resonance peak (Fig. \ref{Fig:incdep}a, inset). The
remaining excitations exhibit 
a strong incident energy dependence, where both antiholon excitations and the
multi-spinon/antiholon excitations centered at $q = 0$ are difficult to resolve
for $\omega_\mathrm{in} \notin (3t, 8t)$. By varying $\omega_\mathrm{in}$, one
selects particular intermediate states $|n\rangle$ in the RIXS process. The
incident energy dependence shown in Fig. \ref{Fig:incdep} indicates that only
certain intermediate states can reach the multi-particle excitations centered
at $q = 0$.   

\begin{figure}[t] 
\centering
\vskip -0.65cm
\includegraphics[width=0.8\linewidth, angle =-90,center]{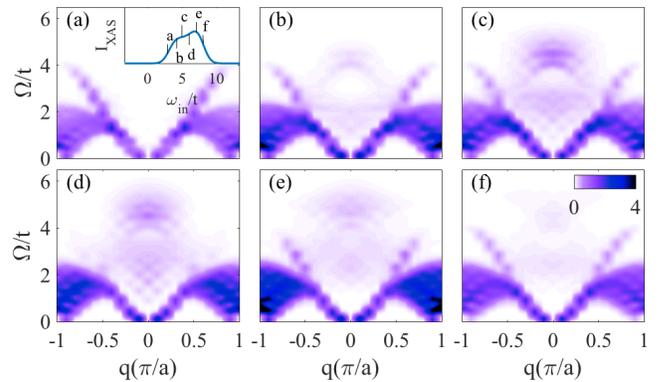}
\vskip -0.75cm
\caption{\label{Fig:incdep} The dependence of RIXS spectra on the incident photon energy $\omega_\mathrm{in}$ for a 5\% doped 20-site chain, evaluated using the full Kramers-Heisenberg formalism. The inset of panel (a) shows the XAS spectrum from the model, along with the incident energies used in each of the RIXS calculations.}
\end{figure}
 
 \textit{Discussion} ---  Several previous theoretical works have
calculated the RIXS spectra for 1D $t$-$J$
\cite{PhysRevB.75.115118,PhysRevLett.106.157205} and Hubbard
\cite{PhysRevB.85.064423,PhysRevB.85.064422} chains using the same formalism. 
In the doped and undoped cases, these studies obtained RIXS spectra resembling
$S(q,\omega)$; however, they did not capture the (anti)holon or multi-spinon
excitations observed here.  
 Refs.~\cite{PhysRevB.85.064423, PhysRevB.75.115118}, obtained nonzero weight
in the $q = 0$ response but with a significantly reduced spectral weight in
comparison to our results. In RIXS at oxygen K-edge, only $\Delta S=0$
excitations are allowed. Ref. \onlinecite{PhysRevB.85.064423,
PhysRevB.85.064422} showed that $\Delta S =0 $ excitations vanishes at
$q=\pi/a$, whereas we have the maximum at that point in our model. We believe
that this discrepancy is due to the 
lack of hopping from the core-hole site due to the 
strong core-hole potential used in that
work, which is appropriate for the Cu $L$ and $K$-edges.  A strong core-hole
potential will tend to localize the excited electrons in the intermediate
state, thus suppressing its dynamics. We can confirm this in our model by
setting $t=0$ in the intermediate state for the undoped system, which also
prohibits charge fluctuations and produces spectra similar to Refs.  
\onlinecite{PhysRevB.85.064423, PhysRevB.85.064422}. 
Furthermore, given the sensitivity to $\omega_\mathrm{in}$ shown in Fig. 4,  
prior studies may have missed the relevant 
excitations due to their choice of incident energies. 

In summary, we have shown that spin-charge separation can be
observed in O $K$-edge RIXS on doped 1D-AFMs and that these systems exhibit
remarkably rich spectra consisting of multi-spinon and holon excitations.  
Our
results highlight the potential for RIXS to simultaneously access the charge,
spin, and orbital degrees of freedom in fractionalized quasiparticle
excitations, applicable to many quantum materials.  

{\it Acknowledgements} --- We thank C. D. Batista, J. Schlappa, T. Schmitt, and
K. Wohlfeld for useful discussions. A. N. and E. D. were supported by the U.S.
Department of Energy, Office of Basic Energy Sciences, Materials Sciences and
Engineering Division. This research used computational resources supported by
the University of Tennessee and Oak Ridge National Laboratory’s Joint Institute
for Computational Sciences.

 \bibliography{1Ddoped.bib}
\end{document}